\documentclass[9pt,twocolumn,twoside]{opticajnl}

\journal{ol} 

\setboolean{shortarticle}{true}

\title{Design and Verification of a Flat-Top Beam Shaper Utilizing Spherical Lenses }

\author[1,2]{Wei Yang}
\author[1,2,*]{Zhen Zhang}

\affil[1]{State Key Laboratory of Tribology in Advanced Equipment, Department of Mechanical Engineering, Tsinghua University, Beijing 100084, China}
\affil[2]{Beijing Key Laboratory of Precision/Ultra-precision Manufacturing Equipments and Control, Tsinghua University, Beijing 100084, China}

\affil[*]{Corresponding author: zzhang@tsinghua.edu.cn}

\begin{abstract}
We propose a novel coaxial spherical-lens shaping system to achieve a flat-top focus profile by exploiting the spherical aberration of the lens. The well-known aspherical-lens shaping system is compared to ours, and a comparison of etching on silicon is conducted between a Gaussian beam and the one shaped by the proposed system. Analytical and experimental results show that the proposed system is able to achieve flat-top beam shaping using only spherical lenses. Compared with the existing systems, our proposed one is capable of shaping Gaussian beams with different wavelengths and waist radii. Furthermore, the shaped one obtained by ours is the flat-top beam with extended depth of focus. This letter provides an efficient and precise flat-top beam shaping scheme for optics and laser technology, and increases the potential of spherical lenses in the field of laser shaping.
\end{abstract}

\setboolean{displaycopyright}{true}

\begin{document}

\maketitle

Flat-top beam is of a uniform cross-section intensity distribution, which prevents excessive local energy in the focal spot that would destroy manufacturing accuracy or material properties comparing to Gaussian beams~\cite{malinauskas2016ultrafast,li2002light}. Due to the superior quality and manufacturing precision, flat-top beams are widely used in various optics and laser techniques~\cite{zhang2022all,chen2019study,tervonen1993acousto,weber2019use}. In selective laser melting (SLM), which is the main technology used in industrial additive manufacturing, the use of flat-top beams can effectively overcome the problems of uneven heating and low-sintering-molding efficiency associated with Gaussian beam manufacturing processes~\cite{wang2020developing,okunkova2014experimental}. In laser drilling, the use of flat-top beams can effectively reduce the heat-affected zone, recast layer thickness and minimize hole taper comparing to Gaussian beams~\cite{flamm2021structured,banerjee2013detection}. However, the output beam of most lasers is of Gaussian optical intensity distribution. There is a demand for beam shaping device to transform the Gaussian beam into a flat-top one.

\begin{figure}[htbp]
	\centering
	\fbox{\includegraphics[width=\linewidth]{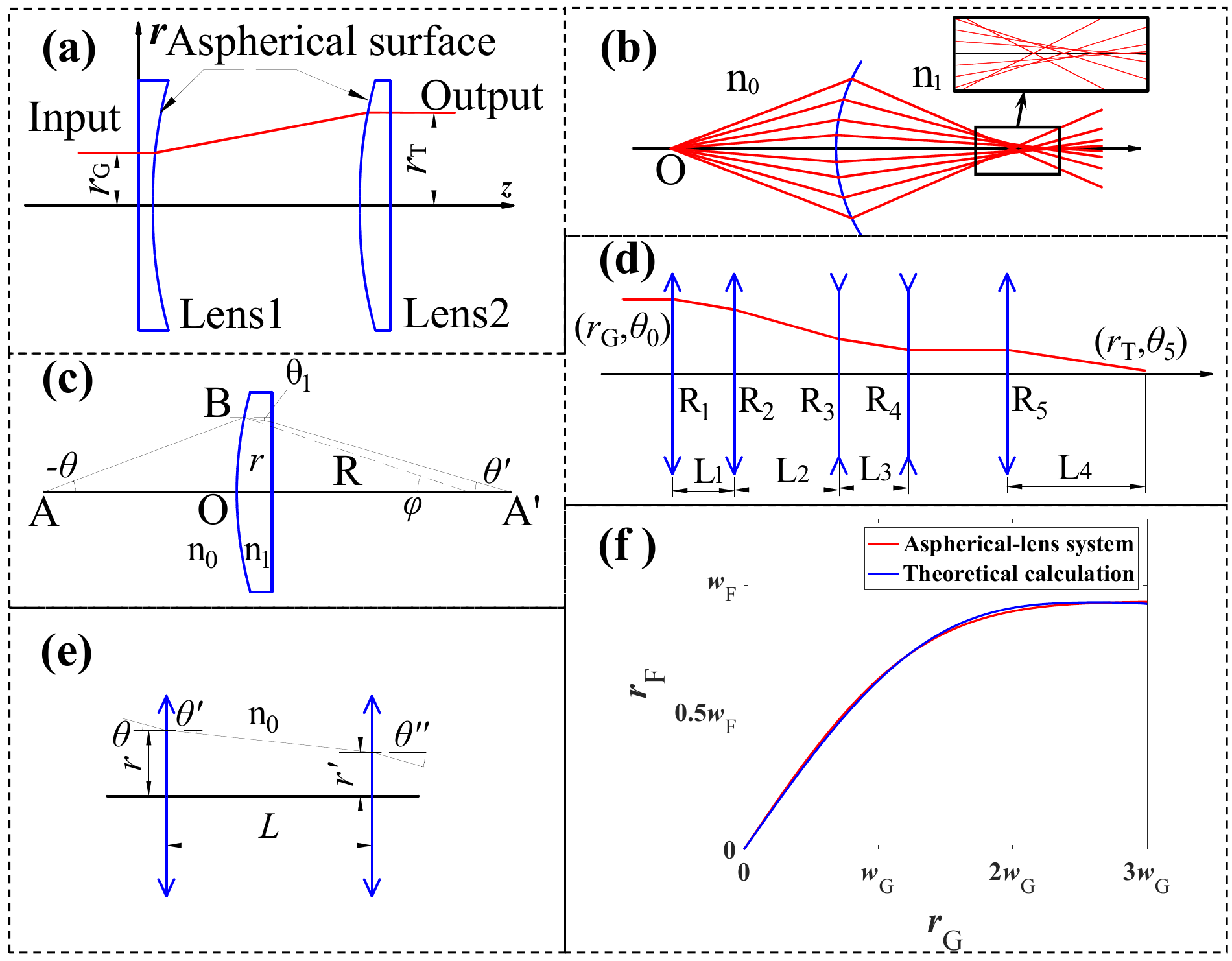}}
	\caption{ (a) Schematic of the aspherical-lens shaping system. (b) Schematic of the spherical lens aberration. (c) Schematic of the correspondence between object and image space aperture angles of spherical lenses. (d) Schematic of the coaxial spherical-lens shaping system. (e) Schematic of rays transmitted between lenses. (f) The coordinate correspondence of the incident and exit rays between the radial coordinates of the Gaussian and flat-top beams.}
	\label{geometric_optics_schematic}
\end{figure}

There are various methods for shaping Gaussian beams into flat-top beams, including aspherical-lens systems~\cite{kasinski1997near,tarallo2007generation}, diffractive optical elements (DOE)~\cite{de2006beam,kopp1999efficient}, liquid-crystal spatial light modulators~\cite{hendriks2012generation,ma2010generation}, metasurfaces, and metamaterials~\cite{li2019wideband,rizza2011two}. Among these methods, the aspherical-lens shaping system offers high shaping efficiency, high energy conversion, and a simple structure, and hence it is widely studied and employed. This system achieves beam shaping by designing an aspheric surface based on the coordinate mapping of Gaussian beams and flat-top beams~\cite{shafer1982gaussian}. Fig.~\ref{geometric_optics_schematic}(a) shows the schematic of the aspherical-lens shaping system, where the incident Gaussian beam has an irradiance distribution of:
\begin{equation}
	E_{G}(r_{G})=E_{0 G} \exp \left(-\frac{r_{G}{ }^{2}}{w_{0}{ }^{2}}\right),
\end{equation}
where $E_{0 G}$ is the complex amplitude on the optical axis, and $w_{0}$ is the waist radius, and $r_{G}$ is the radial coordinate of each ray. The irradiance distribution of the ideal flat-top beam is:
\begin{equation}
	E_{F}(r_{F})=\left\{\begin{array}{cl}
		E_{0 F}, & (r_{F} \leq w_{F}) \\
		0, & (r_{F}>w_{F})
	\end{array}\right.,
\end{equation}
where $E_{0 F}$ is the complex amplitude on the optical axis, and $w_{F}$ is the spot radius, and $r_{F}$ is the radial coordinate of each ray. According to the edge-ray principle and the law of conservation of energy, the radial coordinate correspondence between the incident Gaussian beam and the output flat-hat beam can be expressed as:
\begin{equation}
	r_{F}=w_{F} \cdot \sqrt{\left(1-\exp (-\frac{2 r_{G}^{2}}{w_{0}^{2}})\right)}.
	\label{coordinate_correspondence}
\end{equation}
According to Eq~(\ref{coordinate_correspondence}), the lens shape required for the aspherical-lens can be obtained by using Snell's law. However, it should be noted that the shape of the aspheric surface is dependent of the diameter and wavelength of the incident Gaussian beam, as dictated by the Eq~(\ref{coordinate_correspondence}) and Snell's law. Moreover, the manufacturing error of the aspheric surface shape may significantly affect the quality of the resulting flat-top beam.

In this letter, we propose and validate a coaxial spherical-lens shaping system for achieving flat-top beams, aimed at addressing the aforementioned disadvantages of the aspherical-lens system. Spherical lenses are commonly employed for beam focusing or expansion in various scenarios~\cite{lee2009near,vsarbort2012spherical}, as they have the capability to converge or diverge beams and are readily available as standard components with well-established manufacturing processes. However, as shown in Fig.~\ref{geometric_optics_schematic}(b), the spherical lens has different convergence capabilities for optical rays at different radial positions because the existence of spherical aberration~\cite{holladay2002new}. It is worth noting that spherical lenses with different radii of curvature are combined coaxially to achieve different correspondences between the aperture angles of object and image space. By coaxially combining spherical lenses with different radii of curvature in the manner shown in Fig.~\ref{geometric_optics_schematic}(d), we achieve the same ray coordinate correspondence between the incident and outgoing beams as shown in Eq~(\ref{coordinate_correspondence}). 

Fig.~\ref{geometric_optics_schematic}(c) shows the rays passing through spherical lenses, and the correspondence between the object and image space aperture angles is:
\begin{equation}
	\theta^{\prime}=\arcsin \left[\frac{n_{1}}{n_{0}} \cdot \sin \left(\arcsin \left(\theta_{1}\right)-\arcsin \left(\frac{n_{0}}{n_{1}} \cdot \theta_{1}\right)+\theta\right)\right],
	\label{aperture_angle_relationship}
\end{equation}
where
\begin{equation}
	\theta_{1}=\frac{r \cos (-\theta)}{R}+\sin (-\theta),
\end{equation}
 and $\theta$ and $\theta^{\prime}$ are the object and image space aperture angles, respectively; and $n_{0}$ and $n_{1}$ are the refractive index of the incident medium and the lens material, respectively; and $r$ is the radial coordinate of the incident rays and $R$ is the radius of curvature of the spherical lens. As shown in Fig.~\ref{geometric_optics_schematic}(e), the radial positions correspondence of the rays transmitted between the spherical lenses is:
\begin{equation}
	r^{\prime}=r-L \cdot \tan \theta^{\prime},
	\label{radial_positions_relationship}
\end{equation}
where $r^{\prime}$ is the radial position of the ray on the exit side lens and $L$ is the axial distance of the two lenses. The radial coordinate correspondence ( $r_{g}$ of the incident ray and $r_{t}$ of the emit ray ) in the proposed shaping system can be determined by sequentially applying Eq~(\ref{aperture_angle_relationship}) and Eq~(\ref{radial_positions_relationship}) to each lens of the system. Note that the beam between the last and penultimate lenses is a collimated one ( aperture angle $\theta=0^{\circ}$), which is designed to make the shaping system suitable for laser scanning and laser dynamic focusing scenarios~\cite{yu2023galvanometer}. Parameter optimization was conducted, and when the curvature radius of each lens surface is as shown in Table~\ref{curvature_radius}, the correspondence between the coordinates of the exit and incident rays of the coaxial spherical-lens shaping system is obtained as shown in Fig.~\ref{geometric_optics_schematic}(f). The blue curve in the figure represents the coordinate corresponding curve obtained by Eq~(\ref{coordinate_correspondence}) when the same Gaussian beam waist radius and flat-top beam radius are used as the proposed system. It is observed that the coordinate correspondence obtained by the coaxial spherical-lens shaping system agrees well with the theoretical calculation of the Gaussian and the flat-hat beams. Therefore, the proposed coaxial spherical-lens system is capable of achieving flat-top beam shaping.

\begin{table}[htbp]
	\centering
	\caption{\bf The radius of curvature of each lens surface of the coaxial spherical-lens shaping system}
	\begin{tabular}{ccccc}
		\hline
		$R_{1}$ &$R_{2}$ &$R_{3}$ &$R_{4}$ &$R_{5}$  \\
		\small
		$90.45$mm &\small $155.04$mm &\small $-51.68$mm &\small $-38.76$mm &\small $91.93$mm \\
		\hline
	\end{tabular}
	\label{curvature_radius}
\end{table}

\begin{figure}[htbp]
	\centering
	\fbox{\includegraphics[width=\linewidth]{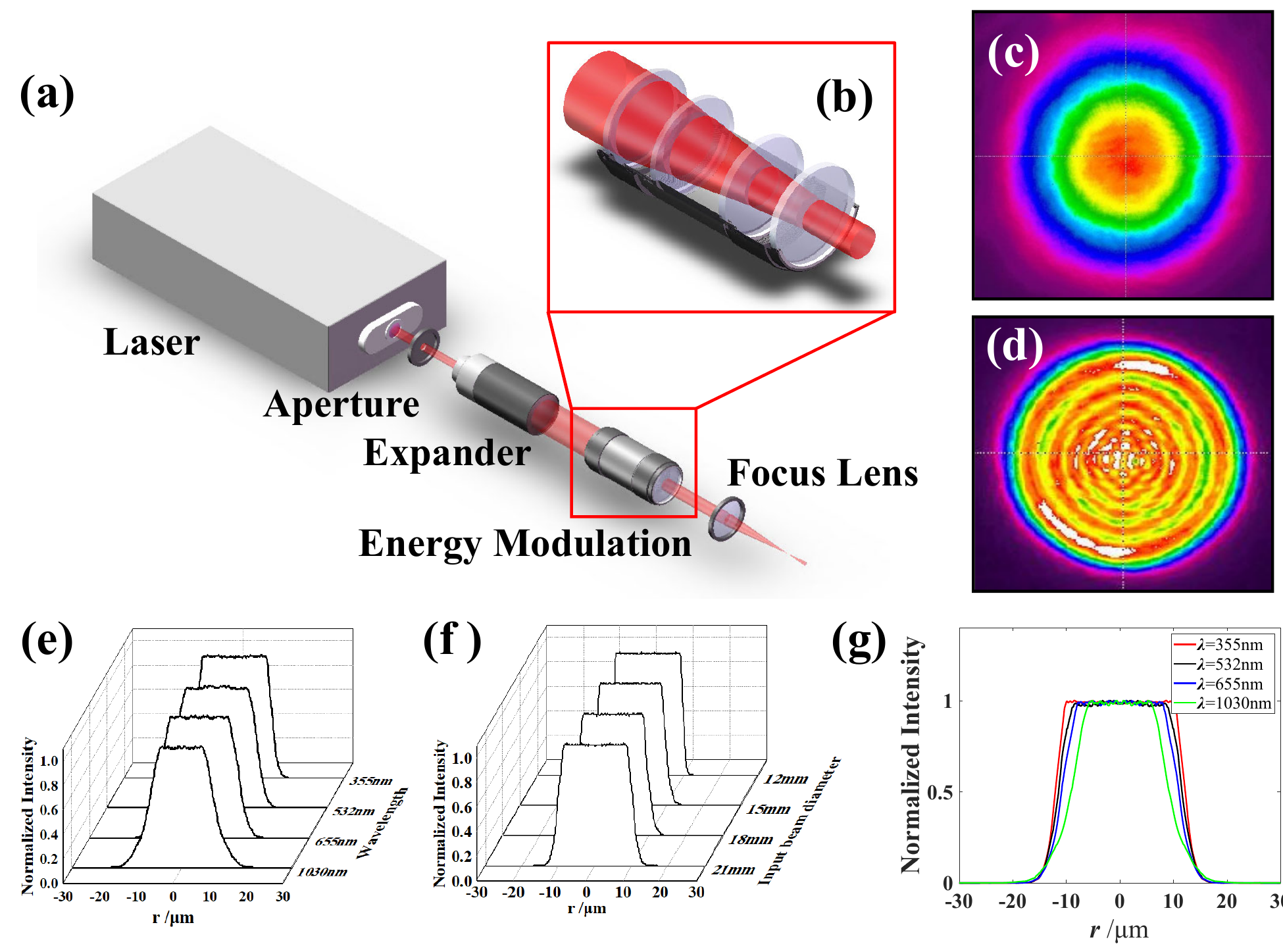}}
	\caption{ (a) Experimental setup of the flat-top beam shaping system. (b) Prototype system of the energy modulation part. (c)-(d) Intensity distributions of the laser beam (c) before and (d) after shaping. (e)-(g) Simulation results of transverse intensity distribution of the spot after beam shaping under (e) different incident laser wavelengths and (f) different waist diameters as well as (g) the distribution comparison.}
	\label{beam_shaping_system}
\end{figure}

In order to validate the flat-top beam shaping performance of the proposed coaxial spherical-lens shaping system, a prototype system is built as depicted in Fig.~\ref{beam_shaping_system}(a). A fiber femtosecond laser (Beilin Laser Axinite IR-05, central wavelength $= 1030$nm, maximum power $= 5$W, pulse duration $<500$fs, and output beam diameter = $1.2\pm0.2$mm) is utilized as the Gaussian beam source. The laser beam is shaped by the aperture and then expanded by the laser beam expander (beam expanding ratio $\beta=15$) to meet the beam waist requirements for flat-top beam shaping. Since the beam between the last and penultimate lenses in the spherical-lens shaping system is collimated, the system is divided into two parts: energy modulation and beam focusing. As illustrated in Fig.~\ref{beam_shaping_system}(b), the first four lenses of the shaping system (DaHeng Optics, GCL-010113, GCL-010155, GCL-010307 and GCL-010306) constitute the energy modulation part, while the last lens (DaHeng Optics, GCL-010814, EFL $= 100$mm) alone constitutes the beam focusing part.

The spot energy distributions before and after flat-top beam shaping are shown in Fig.~\ref{beam_shaping_system}(c) and (d), respectively. It is observed that the beam emitted by the laser exhibits good roundness and the energy distribution is Gaussian. Additionally, the beam spot shaped by the coaxial spherical-lens system displays uniform energy distribution. However, the transverse intensity distribution is oscillating, which deviates from the ideal flat-top beam. This is primarily due to the edge effect of the aperture, resulting in beam diffraction at the edge which affects the transverse intensity distribution of the beam and causes oscillations~\cite{li2018shaping}. The flat-top beam shaping ability of the spherical-lens shaping system for different incident Gaussian beams is analyzed using the physical optics module in the ZEMAX software sequence mode. The beam transverse intensity distribution results of the Gaussian beam with different waist diameters and wavelengths obtained by the proposed system are shown in Fig.~\ref{beam_shaping_system}(e) and (f), respectively. It is observed that better flat-top beam shaping results can be achieved when the waist diameter varies within the range of $12$mm-$21$mm, and the wavelength varies within the range of $355$nm-$1064$nm of the incident Gaussian beam. Notice that the effect of the waist diameter of the incident beam is negligible on the shaped flat-top beam, while the wavelength of the incident beam significantly affects the transverse intensity distribution of the shaped flat-top beam. From the comparison of the transverse intensity distribution of the shaped beam at different wavelengths shown in Fig.~\ref{beam_shaping_system}(g), the full width at half maxima (FWHM) of the shaped flat-top beam is inversely proportional to the wavelength. And the taper of the transverse intensity distribution is proportional to the wavelength of the incident beam. This phenomenon is attributed to the diffraction effect of the beam becoming more prominent as the wavelength increases. Consequently, the coaxial spherical-lens shaping system demonstrates good flat-top shaping ability, and can achieve flat-top shaping of incident beams with different waist diameters and wavelengths using the same system.

\begin{figure}[htbp]
	\centering
	\fbox{\includegraphics[width=\linewidth]{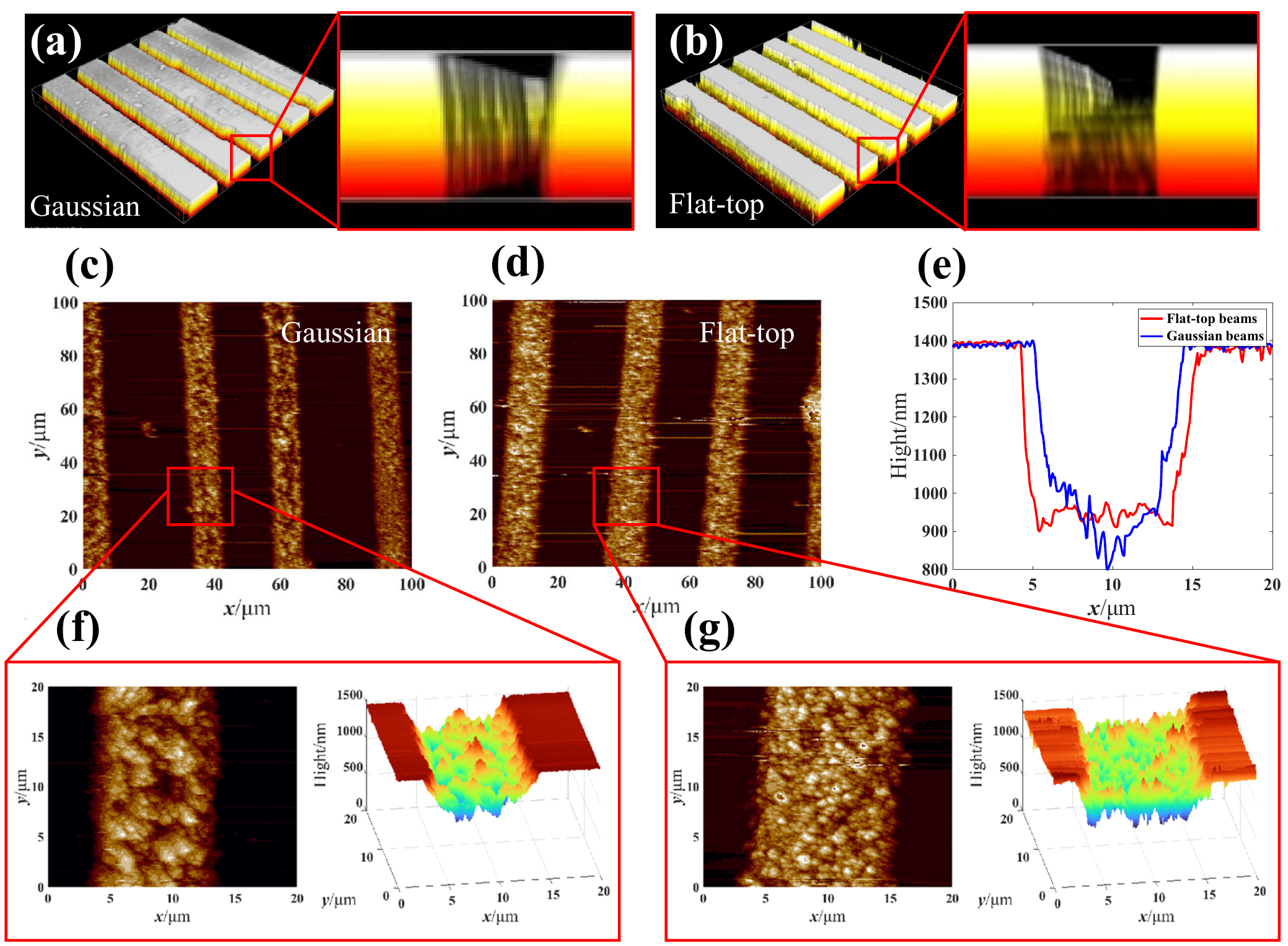}}
	\caption{ (a)-(b) The LSCM images of (a) Gaussian beams and (b) the shaped flat-top beam processing results. (c)-(d) The AFM images of (c) Gaussian beams and (d) the shaped flat-top beam processing results. (e) Comparison of the cross-sectional morphology of grooves processed by Gaussian beams and the shaped flat-top beam. (f)-(g) The AFM images of the partial topography of (f) Gaussian beams and (g) the shaped flat-top beam processing results.}
	\label{process_results_afm}
\end{figure}

In order to evaluate the processing performance of the coaxial spherical-lens shaping system, we conduct laser direct writing of silicon using the femtosecond laser and shaping system mentioned above with Gaussian and flat-top beams. The processed grooves were analyzed using laser scanning confocal microscope (LSCM) and atomic force microscope (AFM) to observe their topographic features. Additionally, the surface topography of the processed substrates is observed and measured using scanning electron microscope (SEM).

A comparison study is conducted on the processing characteristics of Gaussian and the shaped beams for silicon etching under identical incident laser and focusing conditions. Under the conditions of power $=0.84$W, pulse frequency $=496$kHz and focus distance $=100$mm, Fig.~\ref{process_results_afm}(a) and (b) show the LSCM images of the grooved features obtained by processing the substrate with Gaussian and the shaped beams, respectively. It is observed that the tapers of the groove sections obtained using Gaussian and the shaped beams are $5.906^{\circ}$ and $1.216^{\circ}$, respectively. Thus, laser direct writing processing using flat-top beams shaped by the coaxial spherical-lens shaping system resulted in better processing topography taper than Gaussian beams. Fig.~\ref{process_results_afm}(c) and (f) show the AFM images of the grooves and partial topography processed on silicon using Gaussian beam. Correspondingly, Fig.~\ref{process_results_afm}(d) and (g) show the AFM images of the processing results and partial topography using the flat-top beam. Fig.~\ref{process_results_afm}(e) shows the comparison of the cross-sectional morphologies of grooves processed using the Gaussian and the flat-top beams. It is observed that the Gaussian beam processing obtained a groove morphology with a width of $=9.627\mu \mathrm{m}$ and a depth of $=591.8$nm. Correspondingly, the flat-top beam processing resulted in a feature with a width of $=9.627\mu \mathrm{m}$ and a depth of $=591.8$nm. Hence, Gaussian beams can obtain narrower and deeper processing features than those of flat-top beams shaped by the proposed shaping system using the same illuminate beam conditions for laser direct writing processing. However, the cross-sectional shape of the groove obtained by the Gaussian beam is an uneven curve, while the flat-top beam results in a uniform rectangular cross-section.

\begin{figure}[htbp]
	\centering
	\fbox{\includegraphics[width=\linewidth]{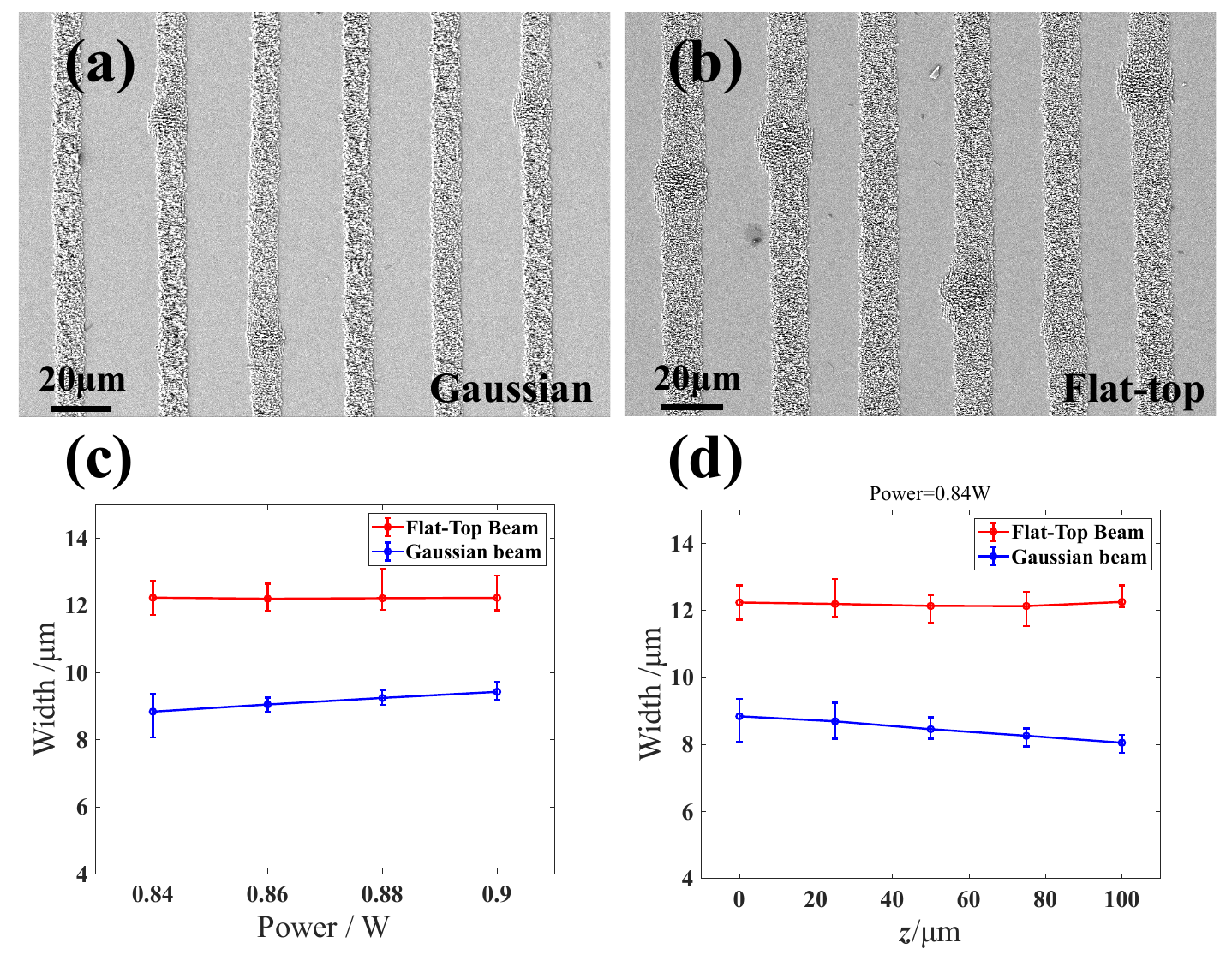}}
	\caption{ (a)-(b) The SEM images of surface topographies of silicon etched by (a) Gaussian and (b) the shaped flat-top beam. (c)-(d) Linewidths of etched features by Gaussian and flat-top beams under different (c)laser power and (d) axial defocusing amount.}
	\label{process_results_sem}
\end{figure}

The morphology obtained by laser direct writing can be influenced by the properties of the illuminating laser and the beam focusing conditions~\cite{meyer2017single,zou2022power,paraschiv2021influence}. SEM is used to accurately obtain the sizes and features of processed topography. Fig.~\ref{process_results_sem}(a) and (b) show the SEM images of the topography obtained by etching silicon with a Gaussian beam and the shaped flat-top beam, respectively. Fig.~\ref{process_results_sem}(c) and (d) show the variations of the obtained etching width with the illumination laser power and the beam defocusing amount, respectively. It is observed that the etching width obtained by the flat-top beam is always larger than that of Gaussian beam under different incident laser power and defocusing amount. The etching width obtained by the Gaussian beam is proportional to the laser power and is inversely proportional to the defocusing amount. While a uniform etching width is obtained by the flat-top beam under different laser power. This is mainly related to the ablation threshold of silicon and the laser energy distribution. The etching width is equal to the diameter of the spot area reaching the ablation threshold of the substrate. The area of the Gaussian beam reaching the substrate ablation threshold is proportional to the laser power. And as the defocusing amount increases, the on-axis optical intensity decreases, which resulting in a smaller spot area that reaches the substrate ablation threshold. Correspondingly, the energy of the flat-top beam is uniformly distributed within the spot, so the area reaching the target ablation threshold does not change with the increase of power. Additionally, the processing width of the flat-top beam fluctuates slightly at $12.186\mu \mathrm{m}$ within the variation range of the defocusing amount of $100\mu \mathrm{m}$, as shown in Fig.~\ref{process_results_sem}(d). This indicates that the flat-top beam obtained by the coaxial spherical-lens shaping system in the focal region can maintain its intensity distribution in a long axial propagation length. It is the feature of extended depth of focus~\cite{pal2018generating,reddy2019robust}. Therefore, the beam shaped by the coaxial spherical-lens shaping system is the flat-top beam with extended depth of focus. 

In conclusion, we have proposed a novel coaxial spherical-lens shaping system for realizing flat-top focus profiles. Analytical and experimental results show that the proposed system successfully shape flat-top beams by employing only spherical lenses. As a standard element with mature technology, spherical lens used in shaping system can obtain higher shaping quality than aspheric lenses. Furthermore, the coaxial spherical-lens system is able to shape Gaussian beams with different wavelengths and waist radii. The etching results of Gaussian and our shaped beams on silicon demonstrate that our shaped one effectively improve the processing shape taper comparing to that of the Gaussian beam. And the flat-top shaping beam can process uniform rectangular etching groove cross-section and maintain a uniform etching width under different laser powers. Additionally, our shaped beam is the flat-top beam with extended depth of focus, as demonstrated by the etching results under different defocusing amounts. Our system provides a cost-effective and highly efficient flat-top beam shaping method for optics laser technology, while also increases the potential of spherical lenses in the field of laser shaping.

\begin{backmatter}
	
	\bmsection{Funding} National Natural Science Foundation of China (No. 52275564).
	
	\bmsection{Disclosures} Competing interest: Z.Z., W. Y., and K. Y. applied for a patent related to this work in China(no.2023101083814).
	
	\bmsection{Data availability} Data underlying the results presented in this paper are not publicly available at this time but may be obtained from the authors upon reasonable request.
	
\end{backmatter}

\bibliography{sample}

\bibliographyfullrefs{sample}


\ifthenelse{\equal{\journalref}{aop}}{%
\section*{Author Biographies}
\begingroup
\setlength\intextsep{0pt}
\begin{minipage}[t][6.3cm][t]{1.0\textwidth} 
  \begin{wrapfigure}{L}{0.25\textwidth}
    \includegraphics[width=0.25\textwidth]{john_smith.eps}
  \end{wrapfigure}
  \noindent
  {\bfseries John Smith} received his BSc (Mathematics) in 2000 from The University of Maryland. His research interests include lasers and optics.
\end{minipage}
\begin{minipage}{1.0\textwidth}
  \begin{wrapfigure}{L}{0.25\textwidth}
    \includegraphics[width=0.25\textwidth]{alice_smith.eps}
  \end{wrapfigure}
  \noindent
  {\bfseries Alice Smith} also received her BSc (Mathematics) in 2000 from The University of Maryland. Her research interests also include lasers and optics.
\end{minipage}
\endgroup
}{}

\end{document}